\documentclass[pra,nofootinbib,twocolumn,showpacs]{revtex4}

\usepackage{amsmath,amssymb,amsthm,graphicx,color}

\newtheorem{Thm}{Theorem}
\newtheorem{Lem}{Lemma}

\newtheorem{Claim}{Claim}
\theoremstyle{definition}
\newtheorem{C}{Counter Example}
\newtheorem*{Proof}{Proof}

\newcommand{\HA}{\mathop{\mathcal{H}}\nolimits}
\newcommand{\B}{\mathop{\mathcal{B}}\nolimits}

\newcommand{\CA}{\mathop{\mathbb{C}}\nolimits}
\newcommand{\N}{\mathop{\mathbb{N}}\nolimits}

\newcommand{\bra}[1]{\langle #1 |}

\newcommand{\ket}[1]{| #1 \rangle}

\newcommand{\bracket}[2]{\langle #1 | #2 \rangle}

\newcommand{\ketbra}[2]{| #1 \rangle \langle #2 |}

\begin{document}

\title{Comments on ``Best conventional solutions to the King's problem"}
\author{Gen Kimura}
\email{gen@ims.is.tohoku.ac.jp}
\author{Hajime Tanaka}
\email{htanaka@ims.is.tohoku.ac.jp}
\author{Masanao Ozawa}
\email{ozawa@math.is.tohoku.ac.jp}
\affiliation{Graduate School of Information Sciences, Tohoku University, Aoba-ku, Sendai 980-8579, Japan}
\date{\today}
\begin{abstract}
Conventional solutions to the (Mean) King's problem without using entanglement have been investigated by Aravind [P. K. Aravind, ``Best conventional solutions to the King's problem'', 
Z. Naturforsch. \textbf{58a}, 682 (2003)].
We report that the upper bound for the success probability claimed there is not valid in general, and give a condition for the claim to be justified.    
\end{abstract}
\pacs{Quantum State Retrodiction, Mean King's Problem, Mutually Unbiased Bases}

\maketitle

The (Mean) King's problem \cite{MKP1,MKP2,MKP3} is a kind of quantum estimation problem with delayed (classical) information and has been studied in detail \cite{MKP2,MKP3}, relating with an unsolved problem on the existence of a maximal set of mutually unbiased bases (MUBs) \cite{MUB}.
The standard approach to solve the King's problem is to utilize entanglement, and nowadays it is shown that the success probability is in fact $1$ for any $d$ level system \cite{MKP3}.
On the other hand, Aravind \cite{Aravind2003ZN} has considered the King's problem without using entanglement, in order to elucidate the role of entanglement in this problem.
In this case he claimed the following upper bound of the success probability $P(d)$ for any $d$ level system: 
$$
	P(d) \le \frac{2\sqrt{d}+d-1}{\sqrt{d}(d+1)}.
$$
For small $d$, this gives
\begin{equation*}
\begin{array}{ccccccc}
	\hline\hline
	d & 2 & 3 & 4 & 5 & 8 & 9 \\
	\hline
	P(d) & 0.9024 & 0.7887 & 0.7000 & 0.6315 & 0.4972 & 0.4667 \\
	\hline
\end{array}
\end{equation*}
The purpose of this letter is, however, to show that this upper bound is not justified for $d\ge 3$ in general. We also give a condition for the claim to be justified.

We begin by recalling the setting of the King's problem without entanglement, as formulated in \cite{Aravind2003ZN} by the following steps (A)-(E)F

(A) A physicist, Alice, prepares a $d$-level quantum system $S$ in a state of her choosing and gives it to the king.

(B) The king carries out a projective measurement with respect to one of $d+1$ MUBs $\{\ket{\Psi^{\mu}_j}\}_{j=1}^d (\mu = 0,\ldots,d)$ \cite{not:MUB}, and notes the output he obtained. 

(C) The physicist carries out a control measurement in an orthonormal basis $\chi=\{\ket{\chi_k}\}_{k=1}^d$ on the system $S$.

(D) The king reveals which of the MUBs he has measured.
(Delayed classical information.)

(E) The physicist is required to correctly predict the output of the king's measurement in Step (B). 

\noindent
Let the physicist prepare a density operator $\rho$ of the system $S$ in Step (A). 
In Step (B), the king is supposed to randomly choose one of $d+1$ MUBs. 
In other words, the probability that the king chooses the $\mu$th MUB is $\frac{1}{d+1}$.
(Therefore, what we consider is a Bayes estimation problem with a uniform prior distribution.)
It is assumed that the king's measurement is a projective measurement with respect to one of given $d+1$ MUBs, so that the post measurement state in Step (B) is $\ket{\Psi^\mu_j}$ if the king chose the $\mu$th MUB and obtained his output $j$.
In Step (C), the physicist is allowed to make measurements only on the system itself (the ``conventional" solution \cite{Aravind2003ZN}) which are described by an orthonormal basis $\chi=\{\ket{\chi_k}\}_{k=1}^d$ (namely a measurement of a non-degenerate observable) \cite{not:entanglement}.
The conditional probability to obtain an output $k$ given $\mu,j$ and a basis measurement $\chi$ is given by $P(k|\mu,j,\chi)=|\bracket{\Psi^\mu_j}{\chi_k}|^2$.  
Finally, in Step (E) the physicist is required to prepare a decision function $s: (k,\mu) \mapsto s_{k\mu} \in \{1,\ldots, d\}$ by which she guesses the king's output $j$ based on her output $k$ obtained in Step (C) and the post information $\mu$ revealed in Step (D).  
(Namely, she predicts the king's output to be $s_{k\mu}$ if she obtained the pair $(k,\mu)$ of her output $k$ and the king's MUB $\mu$.)
Based on this setting, the physicist's strategy is to find a suitable input state $\rho$, a basis measurement $\chi$, and a decision function $s$ which maximize her success probability $P_d(\rho,\chi,s)$ to correctly predict King's output, where it is determined as $P_d(\rho,\chi,s)=\sum_{\mu=0}^d \frac{1}{d+1}P(\rho,\chi,s,\mu)$ with the conditional success probability $P(\rho,\chi,s,\mu)$ given by 
$ P(\rho,\chi,s,\mu) = \sum_{j=1}^d P(j|\rho,\mu) \sum_{k=1}^d\delta_{j,s_{k\mu}} P(k|\mu,j,\chi)$. 
Thus the success probability $P_d(\rho,\chi,s)$ with respect to the physicist's choice $(\rho,\chi,s)$ for a $d$ level system is given by
\begin{multline}\label{eq:prob}
	P_d(\rho,\chi,s) \\
	=\frac{1}{d+1}\sum_{\mu=0}^d\sum_{k=1}^d\bra{\Psi^\mu_{s_{k\mu}}}\rho\ket{\Psi^\mu_{s_{k\mu}}}|\bracket{\Psi^\mu_{s_{k\mu}}}{\chi_k}|^2. 
\end{multline}
Here we have two remarks on $P_d(\rho,\chi,s)$. 
First, since it is an affine function (and hence a convex function) with respect to the input state $\rho$, its maximum is attained by a pure state. 
Second, with fixed $\rho$ and $\chi$, the optimal decision function $s_{\max}$ is given by
\begin{equation}\label{OptimalGuessFunction}
s_{\max}(k,\mu) \equiv \underset{j=1,\ldots,d}{\mathrm{argmax}}\left[ \bra{\Psi^\mu_j} \rho \ket{\Psi^\mu_j} |\bracket{\Psi^\mu_{j}}{\chi_k}|^2\right].
\end{equation}
(Here $\mathrm{argmax}_{j=1,\dots,d} [F(j)]$ assigns a value $j$ which maximizes the real function $F$ on $\{1,\dots,d\}$.)
Indeed, from \eqref{eq:prob} it is easy to see $P_d(\rho,\chi,s)\le P_d(\rho,\chi,s_{\max})$ for any decision function $s$. 

Aravind claimed an upper bound of the success probability \eqref{eq:prob}: 
\begin{Claim}[Aravind \cite{Aravind2003ZN}]\label{The:Aravind}
For any density operator $\rho$, a basis measurement with respect to an orthonormal basis $\chi = \{\ket{\chi_k}\}_{k=1}^d$, and a decision function $s: (k,\mu)\mapsto s_{k\mu} \in\{1,\ldots,d\}$, the success probability \eqref{eq:prob} on the guess of the king's output is bounded from above by
\begin{equation}\label{eq:AravindBound2}
	P_d(\rho,\chi,s)\le P_d(\ketbra{\Psi^\mu_j}{\Psi^\mu_j},\chi,s) \le \frac{2\sqrt{d}+d-1}{\sqrt{d}(d+1)}, 
\end{equation} 
where $j \in \{1,\ldots,d\}$ and $\mu \in \{0,\ldots,d\}$. 
\end{Claim}
Claim \ref{The:Aravind} includes two statements: (i)
\emph{An optimal input state can be taken from one of the basis vectors $\ket{\Psi^\mu_j}$ of the MUBs} (the first inequality in \eqref{eq:AravindBound2}); 
and (ii) it is bounded from above by $\frac{2\sqrt{d}+d-1}{\sqrt{d}(d+1)}$ (the second inequality in \eqref{eq:AravindBound2}). 

\begin{figure} 
\includegraphics[height=0.45\textwidth]{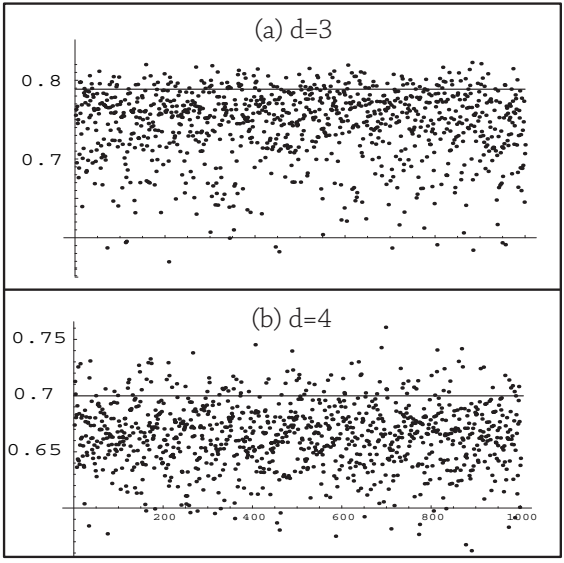}
\caption{
For (a) $d=3$ and (b) $d=4$, the success probability $P_d(\rho,\chi,s_{\max})$ is plotted for $1000$ randomly generated PVM measurements.
The input states are taken from \eqref{eq:input1} for $d=3$ and from \eqref{eq:input2} for $d=4$.
The solid lines stand for Aravind's bound \eqref{eq:AravindBound2}.
One finds several counter examples which surpass his bound.}\label{fig.CE1}  
\end{figure}

Indeed \eqref{eq:AravindBound2} is true when $d=2$, and one can also find a suitable choice of $(\rho,\chi,\mu)$ to attain the bound \cite{SSWEKW2003PRL}. 
However, as we shall see some counter examples, it is not justified for $d \ge 3$ in general. 
\begin{C}[Case $d=3$]
Let $\{\ket{\Psi^\mu_j}\}_{j=1}^3$ $(\mu=0,\ldots,3)$ denote the MUBs constructed by Ivanovi\'{c} \cite{Ivanovic1981JPA}.
Let $\rho=\ketbra{\varphi}{\varphi}$ where 
\begin{equation}\label{eq:input1} 
	\ket{\varphi}=(0,\frac{i}{\sqrt{2}},\frac{3+\sqrt{3}i}{2\sqrt{6}})^T,
\end{equation}
and let $\chi = \{\ket{\chi_k}\}_{k=1}^3$ be the orthonormal basis defined by
\begin{align*}
	\ket{\chi_1} &= (\frac{1}{\sqrt{2}},\frac{\sqrt{3}i}{2\sqrt{2}},- \frac{\exp(\frac{3 i \pi}{4})}{2 \sqrt{2}})^T, \\
\ket{\chi_2} &= ( \frac{i}{\sqrt{2}},{\frac{\sqrt{3}}{2\sqrt{2}}},-\frac{\exp(\frac{i \pi}{4})}{2 \sqrt{2}})^T, \\
\ket{\chi_2} &= (0, \frac{\exp(\frac{-i \pi}{4})}{2},\frac{\sqrt{3}}{2})^T.  
\end{align*}
Then, the success probability $P_3(\rho,\chi,s_{\max} )$ with the optimal guess function \eqref{OptimalGuessFunction} is given by
\begin{equation}\label{eq:c1}
	P_3(\rho,\chi,s_{\max})=\frac{21 + 2\sqrt{2} + 6\sqrt{6}}{32} \simeq 0.8212.
\end{equation}
This is strictly greater than Aravind's bound $\frac{3+\sqrt{3}}{6}\simeq 0.7887$ for $d=3$.
\end{C}
Aravind \cite{Aravind2003ZN} also showed that the bound \eqref{eq:AravindBound2} is attainable for $d=4$.
To see that, he used the following MUBs: 
\begin{equation*}
\begin{array}{cccc}
\hline
\ket{\Psi^0_1}=1000 & \ket{\Psi^0_2}=0100 & \ket{\Psi^0_3}= 0010 & \ket{\Psi^0_4} = 0001 \\
\ket{\Psi^1_1}=1111 & \ket{\Psi^1_2} = 1\bar{1}1\bar{1} & \ket{\Psi^1_3}= 11\bar{1}\bar{1}& \ket{\Psi^1_4}= 1\bar{1}\bar{1}1 \\
\ket{\Psi^2_1}= 1ii\bar{1}& \ket{\Psi^1_2}= 1\bar{i}i1& \ket{\Psi^1_3}= 1i\bar{i}1& \ket{\Psi^1_4}= 1\bar{i}\bar{i}\bar{1} \\
\ket{\Psi^3_1}= 1\bar{1}ii& \ket{\Psi^3_2}= 11\bar{i}i& \ket{\Psi^3_3}= 11i\bar{i}& \ket{\Psi^3_4} = 1\bar{1} \bar{i}\bar{i} \\
\ket{\Psi^4_1} = 1i\bar{1}i& \ket{\Psi^4_2}= 1\bar{i}1i& \ket{\Psi^4_3} = 1i1\bar{i}& \ket{\Psi^4_4} = 1\bar{i}\bar{1} \bar{i} \\
\hline
\end{array}
\end{equation*}
(The shorthand notation $\ket{\Psi^\mu_j}=abcd$ indicates that $\ket{\Psi^\mu_j}$ has the (unnormalized) form
$(a,b,c,d)^T \in \CA^4$, and $\bar{a}$ stands for the negative of $a$.)
However, we can also construct the following counter example in these MUBs:
 
\begin{C}[Case $d=4$]
Let $\{\ket{\Psi^\mu_j}\}_{j=1}^4$ $(\mu=0,\ldots,4)$ be the MUBs given in the above table.
Let $\rho=\ketbra{\varphi}{\varphi}$ where 
\begin{equation}\label{eq:input2}
	\ket{\varphi}=\frac{1}{\sqrt{2}}\left(1, 0, -1,0\right)^T,
\end{equation}
and let $\chi=\{\ket{\chi_k}\}_{k=1}^4$ be the orthonormal basis defined by  
\begin{align*}
	\ket{\chi_1}&=\left(\frac{\sqrt{3}i}{2},\frac{9+3\sqrt{3}i}{32},\frac{-\sqrt{3}+i}{32},\frac{3-3\sqrt{3}i}{16}\right)^T, \\
	\ket{\chi_2}&=\left(\frac{\sqrt{3}+i}{4},\frac{-9\sqrt{3} + 9i}{32},-\frac{\sqrt{3}i}{16},\frac{3\sqrt{3} + 9i}{16}\right)^T, \\ 
	\ket{\chi_3}&=\left(0,\frac{5-5\sqrt{3}i}{16},\frac{-3\sqrt{3}-9i}{16},\frac{3+\sqrt{3}i}{8}\right)^T, \\
	\ket{\chi_4}&=\left(0,\frac{-3+\sqrt{3}i}{8},-\frac{3i}{4},\frac{-1-\sqrt{3}i}{4}\right)^T.
\end{align*}
Then the success probability is given by 
\begin{equation}\label{eq:c2}
P_4(\rho,\chi,s_{\max} ) = \frac{6493 + 1065\sqrt{3}}{10240} \simeq 0.8142,
\end{equation}
which is also strictly greater than Aravind's bound $0.7$ for $d=4$. See also Fig.~\ref{fig.CE1} for numerically generated counter examples.
\end{C}

\noindent
Since the computations to obtain \eqref{eq:c1} and \eqref{eq:c2} are just elementary, we only comment on our choices of the input states \eqref{eq:input1} and \eqref{eq:input2}.
Since the optimal success probability can always be attained by pure input states as mentioned before, we may assume $\rho=\ketbra{\varphi}{\varphi}$ for a unit vector $\ket{\varphi}$ without loss of generality.
Now suppose we have $|\bracket{\Psi^{\mu_0}_{j_0}}{\varphi}|^2=0$ for some pair $(j_0,\mu_0)$.
In such a case, if the king chose $\mu_0$ as his measurement basis, then it is already guaranteed (without resort to the physicist's measurement result) that the king's output is absolutely distinct from $j_0$. 
This suggests that a better strategy of the physicist is to find an input state which has as many pairs $(j,\mu)$ satisfying $|\bracket{\Psi^\mu_j}{\varphi}|^2=0$ as possible.
In fact, our input states \eqref{eq:input1} and \eqref{eq:input2} satisfy the condition $|\bracket{\Psi^\mu_j}{\varphi}|^2=0$ at $(j,\mu)=(2,1),(1,2),(1,3),(1,4)$ and $(j,\mu) = (2,0),(4,0),(1,1),(2,1),(2,4),(3,4)$, respectively.

Now we point out a gap in the proof in \cite{Aravind2003ZN}.
Indeed, it is the first inequality in Claim \ref{The:Aravind} that cannot be justified, while the second inequality is verified to be true. 
The crucial error lies in the choice of the decision function $s$ in \cite{Aravind2003ZN}.
Namely, Aravind considered essentially only the decision function given by $s(k,\mu)=f_\mu^{-1}(k)$ where $f_\mu(j) \equiv \underset{k=1,\ldots,d}{\mathrm{argmax}} \left[\bra{\Psi^\mu_j} \rho \ket{\Psi^\mu_j} |\bracket{\Psi^\mu_{j}}{\chi_k}|^2 \right]$ provided that $f_\mu(j)$ is bijective for each $\mu$.
In particular, this restriction forces that $s$ must also be bijective for each $\mu$.
However this decision function is not optimal in general, and moreover
if we choose the input state $\rho$ as discussed above, then the optimal decision function $s_{\max}$ given in \eqref{OptimalGuessFunction} is not injective for many $\mu$.
Based on this false inference, an argument to ``prove" the first inequality is given in \cite[Appendix]{Aravind2003ZN}.

We remark that the second inequality in Claim \ref{The:Aravind} is still valid.
Thus Aravind's result can now be stated in the following weaker form: 
\begin{Thm}\label{Thm:A2}
Let $\{\ket{\Psi^\mu_j}\}_{j=1}^d$ $(\mu=0,\ldots,d)$ be a set of $d+1$ MUBs.
If the input state $\rho$ is of the form $\rho=\ketbra{\Psi^\mu_j}{\Psi^\mu_j}$, then the success probability $P_d(\rho,\chi,s)$ with respect to any basis measurement $\chi$ and a decision function $s$ is bounded from above as  
\begin{equation}\label{eq:AravindBound3}
P_d(\rho,\chi,s) \le \frac{2\sqrt{d} + d - 1}{\sqrt{d}(d+1)}.
\end{equation}   
\end{Thm}
Aravind obtained this result by the method of Lagrange multipliers, but here we give an alternate (and possibly simpler but more rigorous) proof for the reader's convenience.
We begin with the following lemma:  
\begin{Lem}\label{prop:BF}
Let $L \in \B(\HA)$ be any bounded operator on a Hilbert space $\HA$ of the form
\begin{equation*}
	L = \sum_{i=1}^m \ketbra{\phi_i}{\phi_i},
\end{equation*}
where $m\in\N$ and $\ket{\phi_i} \in \HA$ $(||\phi_i||=1)$.
Then the operator norm $||L|| \equiv \sup_{\ket{\psi} \in \HA_d \ (||\psi||=1)} ||L\ket{\psi}||$ has the following upper bound:
\begin{equation}\label{eq:BoundFormula}
	||L|| \le \lim_{n\to \infty}  \left(\sum_{i_1,\ldots,i_{n}=1}^m \prod_{k=1}^{n-1}|\bracket{\phi_{i_{k}}}{\phi_{i_{k+1}}}|\right)^{\frac{1}{n}}.
\end{equation}
\end{Lem} 
\begin{Proof}
For any $\ket{\psi} \in \HA$, we observe 
\begin{widetext}
\begin{align*}
	||L^n \ket{\psi}||^2 &= \sum_{i_1,j_1,\ldots,i_n,j_n=1}^m \bracket{\psi}{\phi_{i_n}}\left(\prod_{k=1}^{n-1}\bracket{\phi_{i_{k+1}}}{\phi_{i_{k}}}\right) \bracket{\phi_{i_1}}{\phi_{j_1}}\left(\prod_{l=1}^{n-1}\bracket{\phi_{j_{l}}}{\phi_{j_{l+1}}}\right)\bracket{\phi_{j_n}}{\psi} \\
	&\le \left( \sum_{i_1,j_1,\ldots,i_n,j_n=1}^m |\bracket{\phi_{i_1}}{\phi_{j_1}}|\left(\prod_{k=1}^{n-1}|\bracket{\phi_{i_{k}}}{\phi_{i_{k+1}}}|\right)\left(\prod_{l=1}^{n-1}|\bracket{\phi_{j_{l}}}{\phi_{j_{l+1}}}|\right)\right) ||\psi||^2 \\
	&= \left(\sum_{i_1,\ldots,i_{2n}=1}^m \prod_{k=1}^{2n-1}|\bracket{\phi_{i_{k}}}{\phi_{i_{k+1}}}|\right)||\psi||^2
\end{align*}
\end{widetext}
by the Schwarz inequality.
Hence we have 
\begin{equation*}
	||L^n|| \le\sqrt{\sum_{i_1,\ldots,i_{2n}=1}^m \prod_{k=1}^{2n-1}|\bracket{\phi_{i_{k}}}{\phi_{i_{k+1}}}|}.
\end{equation*}
By combining this with the Gelfand formula $||L|| = \lim_{n\to \infty} ||L^n||^{\frac{1}{n}}$, we obtain the bound \eqref{eq:BoundFormula}.
\hfill{$\blacksquare$}
\end{Proof}

\noindent
\textbf{Proof of Theorem \ref{Thm:A2}.}
We may set $\rho=\ket{\Psi^0_1}\bra{\Psi^0_1}$ without loss of generality.
Then
\begin{align}\label{eq:appendix:prob}
	P_d&(\rho,\chi,s_{\max}) \notag \\
	=& \frac{1}{d+1}\left[1 + \sum_{\mu=1}^d\sum_{k=1}^d \frac{1}{d}|\bracket{\Psi^\mu_{s_{\max}(k,\mu)}}{\chi_k}|^2\right] \notag \\
	\le& \frac{1}{d+1}\left[1 + \frac{1}{d} \sum_{k=1}^d \left(\sup_{\ket{\psi}} \left[\sum_{\mu=1}^d|\bracket{\Psi^\mu_{s_{\max}(k,\mu)}}{\psi}|^2\right]\right)\right],
\end{align}
where the supremum is over $\ket{\psi}\in \HA_d$ such that $||\psi||=1$.
Let $L \equiv \sum_{\mu=1}^d \ketbra{\Psi^\mu_{s(\mu)}}{\Psi^\mu_{s(\mu)}}$, where $s$ is any real function on $\{1,\ldots,d\}$.
Since $L$ is positive, it follows that
\begin{equation}\label{eq:appendix:norm}
	||L|| = \sup_{\ket{\psi}} \left[\bra{\psi}L\ket{\psi}\right] = \sup_{\ket{\psi}} \left[\sum_{\mu=1}^d |\bracket{\Psi^\mu_{s(\mu)}}{\psi}|^2 \right],
\end{equation}
where the supremums are over $\ket{\psi} \in \HA_d$ with $||\psi||=1$.
From Lemma \ref{prop:BF} and the property \cite{MUB} of MUBs $|\bracket{\Psi^\mu_i}{\Psi^\nu_j}|^2=\delta_{\mu \nu}\delta_{ij}+ (1-\delta_{\mu \nu})\frac{1}{d}$, we have 
\begin{align}\label{eq:appendix:bound}
	||L|| &\le \lim_{n\to \infty}  \left(\sum_{\mu_1,\ldots,\mu_{n}=1}^d \prod_{l=1}^{n-1}|\bracket{\Psi^{\mu_l}_{s(\mu_l)}}{\Psi^{\mu_{l+1}}_{s(\mu_{l+1})}}|\right)^{\frac{1}{n}} \notag \\
	&= \lim_{n\to\infty} \left( \frac{d\sqrt{d}}{\sqrt{d}+d-1} \right)^{\frac{1}{n}} \frac{\sqrt{d}+d-1}{\sqrt{d}} \notag \\
	&= \frac{\sqrt{d}+d-1}{\sqrt{d}}. 
\end{align} 
From \eqref{eq:appendix:prob}, \eqref{eq:appendix:norm} and \eqref{eq:appendix:bound}, we obtain \eqref{eq:AravindBound3}. \hfill{$\blacksquare$}

To summarize, we have constructed some counter examples to Aravind's general bound \eqref{eq:AravindBound2} in \cite{Aravind2003ZN}.
However, we have reconfirmed that it can be justified with restricted input states (Theorem \ref{Thm:A2}).
We shall investigate the correct bound which is valid for arbitrary input states in the near future.

This work is supported by Grant-in-Aid for JSPS Research Fellows and the SCOPE project of the MIC of Japan.


\begin{thebibliography}{10}
\bibitem{not:MUB}
Here we assume $d$ is such that a maximal set of $d+1$ MUBs exists.
\bibitem{not:entanglement}
If she is allowed to make measurements on the system as well as any ancillary systems which may have been entangled in the preparation phase (Step A), she has a perfect solution proposed recently \cite{MKP1,MKP2,MKP3}.
\bibitem{Ivanovic1981JPA}
Suppose $d$ is an odd prime and let $\{\ket{e_k}\}_{k=1}^d$ denote the standard orthonormal basis of $\CA^d$.
Then, Ivanovi\'{c} \cite {MUB} constructed $d+1$ MUBs by $\ket{\Psi^\mu_j} \equiv \sum_{k=1}^d U^{(\mu)}_{jk}\ket{e_k}$, where $U^{(\mu)}_{jk} = \frac{1}{\sqrt{d}} \exp\left[\frac{2 \pi i}{d} (\mu+1) (j+k-1)^2 \right] \ (\mu= 0,\ldots,d-2)$, $U^{(d-1)}_{jk} = \frac{1}{\sqrt{d}} \exp\left[\frac{2 \pi i}{d} jk \right]$, and $U^{(d)}_{jk} = \delta_{jk}$.
\end{thebibliography}
\end{document}